# Revisiting the semi-flexible entangled chains of polymer in the carbyne model

C.H.Wong*, L.Xue, E.A.Buntov, A.F.Zatsepin

*Institute of Physics and Technology, Ural Federal University, Yekaterinburg, Russia*

*Email:ch.kh.vong@urfu.ru*

Abstract:

The Monte Carlo carbyne model is modified to investigate the glass transition of the semi-flexible entangled polymer chains. The stochastic bombardment between monomers are monitored by Metropolis algorithm with help of the consideration of hard potential while the mobility of monomers is governed by its mass, scattering rate and temperature. Our model is capable to show that the glass transition temperature reduces with decreasing film thickness and the formation of critical voids in the thinner polymer contributing to the glass transition that is much easier than the bulk polymer.

1. Introduction

Glass is one of the most useful materials in daily life such as the window, food containers, camera lens…etc. There are different types of calculations about glass transition. Molecular dynamics simulation is one of the famous tools to determine the glass transition temperature ($T_g$) of amorphous polymers [1-5]. Different characters such as the density of the materials and mean square displacements (MSDs) were calculated to estimate the glass transition temperatures ($T_g$) of the materials [1-5]. Another method to calculate the $T_g$ is to consider the change in thermal expansion coefficients in ensembles held at constant pressure and temperature or the change in the temperature dependence of diffusivity of atoms in ensembles held at constant volume and temperature. For example, the molecular dynamic simulation is capable to tackle the thermo-mechanical properties of the cured epoxy network composed of diglycidyl ether bisphenol A (DGEBA) epoxy resin and tetrahydrophthalic anhy-dride (THPA) curing agent and their single-walled carbon nanotubes (SWCNT) reinforced the epoxy matrix composites where the theoretical $T_g$ is 365 K and 423 K respectively [6]. On the other hand, the coarse grain model and Monte Carlo sampling formalism are proposed to simulate the self-assembly in block copolymers and nanoparticle−copolymer composites [7]. Their approaches are based on the particle-based representation that does not invoke a saddle point approximation but allow implementing the treatment of the large 3D systems [7]. It is used to examine the directed assembly of copolymer blends and nanoparticles on nanopatterned substrates where the simulation results are in good agreement with experiment [7] In addition, D. P. Landau et al proposed a methodology [8] to simulate a semi-flexible polymer chain. However, if there is more than one semi-flexible chain in the system, the entanglements play an important role to the glass transition temperature and therefore we are going to test whether the newly developed Monte Carlo carbyne model [9] can simulate the $T_g$ of semi-flexible multi-chains of polymer or not. The $T_g$ of our polymer system is estimated by plotting the volume as a function of temperature where $T_g$ is defined by finding the point of intersection [10,11] for two extrapolated straight lines in the frozen glassy (below $T_g$) and soften states (above $T_g$) respectively.

The hard potential will be added to the existing carbyne model to study how the $T_g$ reacts with the hardness of monomers. As the mobility of monomers is related to the bombardment between monomers and mean free path, the monomer diffusion should give a strong impact to glass transition. Hence we will proceed to the next step by studying the $T_g$ as a function of the mass of monomers. Through changing the dimension gradually from 3D to 2D in the modified carbyne model, the $T_g$ in different thicknesses will be compared in the third stage. An experimental study of densification and depression in glass transition temperature [12] in polystyrene thin films reported that the increase of free volume in bulk polymer is relatively slower than the thin film. This phenomenon will be interpreted by the amount of critical voids based on the modification of the carbyne model.

2. Monte Carlo Model

The box size of 1000 x 1000 x 1000 contains 6 polymer chains where each chain is made up of 100 spherical monomers. The original carbyne model [9] is slightly modified in the following. The initial condition sets the 600 monomers to occupy within the size of 5 x 5 x 5 in random manner. All monomers are identical with the diameter of 0.8 while the preferred chain length between two consecutive monomers $r_o$ is 1. The bombardment between any monomer in the box is checked by monitoring their separation as the diameter of monomer is given. The scattering cross section of monomers is assumed to be a constant. In the carbyne model, the types of covalent bond between the adjacent carbon atoms depends on Boltzmann excitation and the control of minimal energy. To make the carbyne model compatible to polymer, the chemical bond between the nearest monomers depends on the following interactions only.

$$E_{length} = J_L(r-r_0)^2 \qquad E_{angle} = J_A(\cos(\theta)+1)^2$$

$$E_{vdw} = \begin{cases} J_{vdw}\left(3(r-2)^2 - (r-3)^3 - 1\right) & (0 < r < 3) \\ 0 & (otherwise) \end{cases}$$

The monomers are influencing each other via the approximated Lennard Jones potential $E_{vdw}$ [8] and elastic linear bond energy ($E_{length}, E_{angle}$) in addition to the hard potential per collision $E_{hard}$. For the sake of simplicity, both $J_L$ and $J_A$ equals to 1 while the $J_{vdw}$ is 0.03 in the simulation. The interaction terms are put to the Hamiltonian below to rearrange the alignment of the monomers during the Metropolis steps of 300000. The $i$ refers to the number of monomers along the chain and the $j$ corresponds to the number of chains

$$E = \sum_j\sum_i E_{length}(j,i) + \sum_j\sum_i E_{angle}(j,i) + \sum_j\sum_i E_{vdw}(j,i) + \sum_j\sum_i E_{hard}$$

Our model focuses on simulating the volume enclosed by all the monomers vs temperature upon heating. The simulation starts choosing the chain arbitrarily. In the next step we choose the monomer randomly on the selected chain. During the thermal expansion, the bond length energy becomes more positive in spite of the existence in the approximated Lennard-Jones potential. The bond angle is thus readjusted where three consecutive monomers form a pivot [9]. However, the monomers may collide on each other to weaken the effect of volume expansion through the hard potential in the unit of 2 (unless we specify) per collision. If there is a large hard potential energy, a bigger resistance in volume expansion is eventually formed so that the overall energy of the energetic favorable configuration at equilibrium becomes less positive. The chosen monomer at each step is attempted to move in a distance which is the product of velocity and travel time. This travel distance has to be further multiplied by a random number in order to make the calculation more realistic. The velocity is proportional to the square root of temperature divided by mass. The travel time or scattering time is assumed to be temperature independent for simplicity. After the trial distance is obtained, the assigned monomer will move in a random direction. If the new coordinate of monomer produces a more negative energy state, the trial distance and angle of the monomer is accepted. Otherwise, it returns to the original coordinate. The coordinate of the monomer can also be excited by Boltzmann exponential function simultaneously. The initial condition is regenerated at each temperature. Assume no interaction between the polymer film and substrate, the glass transition temperatures changing from 3D to 2D are controlled by the reduction of the box size along z-axis while the box area along XY plane remains unchanged. Each data in the figure 1-4 are obtained by averaging five Monte Carlo results to enhance the reliability.

3. Results and discussion

Figure 1 studies the glass transition in different hardness of monomers. The glass transition temperature is higher in presence of the stiffer monomer. The role of hard potential to the glass transition is confirmed. The collision between the monomers makes the volume expansion more difficult in presence of the tougher hard potential because a more powerful obstacle needs to overcome. Therefore the glass transition temperature is raised.

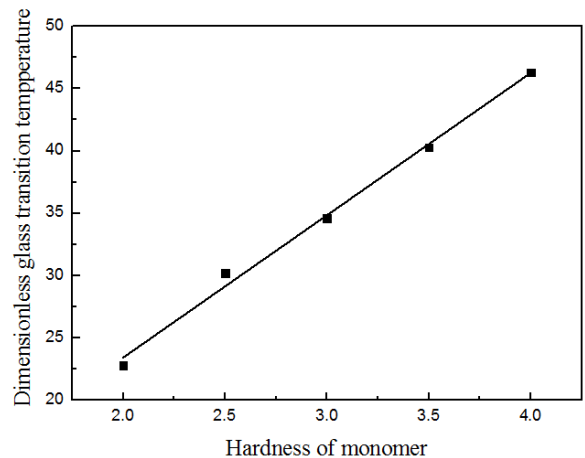

Figure 1: The dimensionless glass transition temperature increases with the hardness of monomers. The mass of monomer and scattering time are scaled to be 10 and 1 respectively.

The volumes as a function of temperature in different masses are presented in figure 2. The heavier monomer it has, the higher glass transition temperature it forms. As the temperature dependence of travel distance of the monomers is considered as concave downward according to Kinect theory, the concave upward-like structure in the figure 2 should involve the collective interaction between the six polymer chains. The sudden change in monomer diffusion rates mark the boundary between glassy and soften state. The polymer with the scaled mass of 10, 15 and 20 show the dimensionless glass transition temperatures T at 22.4, 28.1 and 33.7 respectively. The polymer chains are entangled up so that the volume vs temperature in the figure 2 does not follow the concave downward function owing to the restricted movement of the monomers. As a result, the concave upward-like function appears in the figure 2. Having a higher mobility of monomers is beneficial to get rid of the frozen glassy condition easier upon heating. As a result, a lower glass transition temperature occurs in the lighter monomers in figure 2. It likely agrees with the Flory–Fox equation.

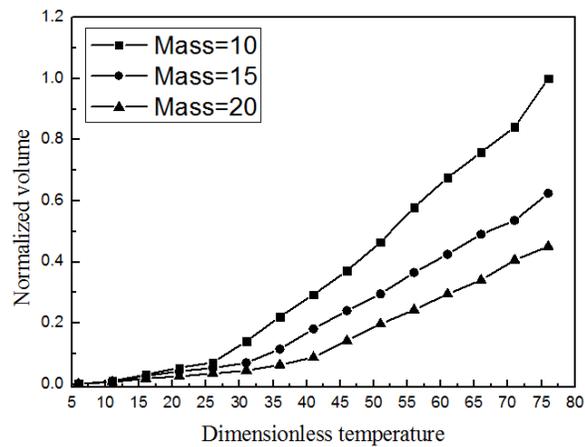

Figure 2: The glass transition occurs at higher temperatures in the heavier monomers. Each curve is fitted by two straight lines in the glassy and soften states. The scattering time of the monomer is scaled to be 1.

The dimensional crossover from 3D to 2D gives a strong impact to glass transition temperature [13]. The reduction of the glass transition temperature is observed in a thinner polymer film as shown in figure 3. The length of box along x and y directions are $L_x = 1000$ and $L_y = 1000$. The glass transition is compared at various thicknesses along z direction. The glass transition temperature, fluctuates from 22.2 to 22.9 only, are obtained when $L_z > 15$. However, the glass transition temperature definitely drops to 8.4 at $L_z = 5$. Another importance of the figure 3 is to allow us to check the thermal expansion of the polymer films below the glass transition temperature in order to study the critical voids [12]. The chains are being heated from $T = 1$. As a consequence of the figure 3, the minimum thickness in our model is $L_z = 5$ and therefore it is safe to study the thermal expansion of the glass films at $T = 6$. The increase of volume in the glassy polymers at $T = 6$ are simulated as illustrated in figure 4. The volume expansion is suddenly enlarged when $L_z$ is smaller than 15. Here the increase of volume $|dV/dT|_{T=6}$ at each thickness is estimated by central differentiation numerically. The reduction of the glass transition temperature in the thinner polymers in figure 3 can be explained with help of the concept of critical void [12] and figure 4.

According to figure 4, the thermal expansion in the thinner polymer films is larger and therefore generating the critical void [9] is relatively easier than the case of thicker film. In other word, more voids give a bigger space to monomer diffusion upon heating to lower the glass transition temperature. The data in figure 3 & 4 are consistent with the experiments [12].

The theoretical elastic modulus of the single carbyne chain is incredibly strong [14,15,16]. By borrowing the concept of the enhanced stiffness of the entangled polymers, we expect that the elastic modulus of the entangled carbyne chains may be much higher than the isolated carbyne. In the lack of computational model to calculate mechanical properties of the entangled chains of carbyne, it is necessary to include the hard potential to the original carbyne model to estimate the elastic modulus of the entangled carbyne chains as a future work.

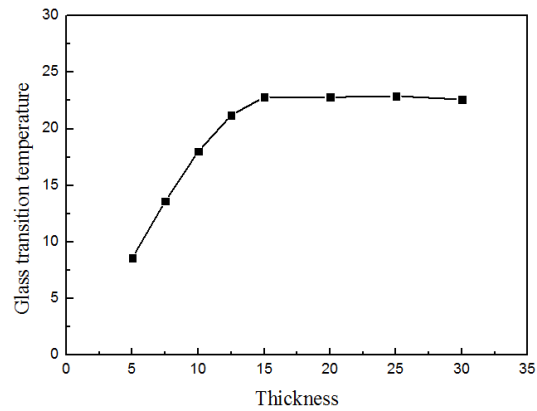

Figure 3: The reduction of the dimensionless glass transition temperature in various thicknesses initiates at $L_z = 15$ where $L_x = L_y = 1000$. The mass and scattering rate of the monomers are set to be 10 and 1 respectively. The data at each thickness is averaged 5 times to minimize the random errors. As a reference, the glass transition temperature equals to 22.4 for $L_x = L_y = L_z = 1000$.

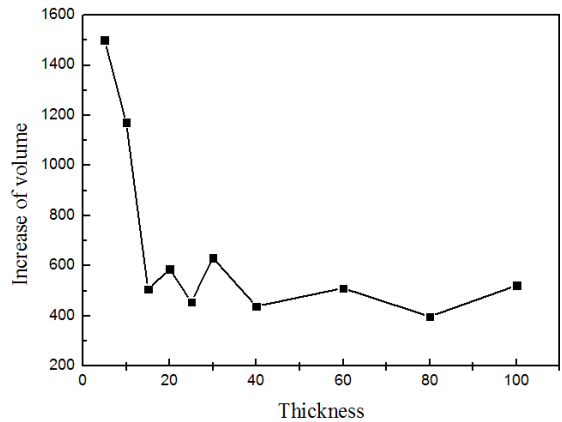

Figure 4: The thermal expansion (dV/dT) of the polymer films as a function of thickness at T = 6. As a reference, the increase of volume of the polymer of $L_x = L_y = L_z = 1000$ equals to 435 at T = 6.

4. Conclusion

We have verified that the carbyne model can be modified to study the glass transition of the multi-chains of polymer. The techniques employed in this Monte Carlo simulation gain more insights on glass transitions. The modified carbyne model utilizing the hard potential to makes the simulation of the entangled polymer chains possible. Both the hardness and molecular weight of monomers play an important role to glass transition. Nevertheless, the glass transition temperature and the rate of thermal expansion can be tuned by the size effect of polymers. The successful prediction of the entangled polymers provided by the modified carbyne model will be the first step for us to simulate the giant elastic modulus of the entangled carbyne chains.